\documentclass{aa}  

\usepackage{graphicx}
\usepackage{amsmath}	
\usepackage{dsfont}

\usepackage{graphicx}
\usepackage{txfonts}
\usepackage[pdfencoding=auto,psdextra]{hyperref}
\hypersetup{
    colorlinks=true,
    linkcolor=blue,
    filecolor=magenta,      
    urlcolor=blue,
    citecolor=blue
}
\usepackage[nameinlink,capitalise]{cleveref}
\usepackage[dvipsnames,table,xcdraw]{xcolor}

\begin{document}

   \title{Probabilistic Estimators of Lagrangian Shape Biases: Universal Relations and Physical Insights}

   \author{F. Maion
          \inst{1,2}
      \and
          J. St\"ucker\inst{1,3}
      \and
          R. E. Angulo\inst{1,4}
          }

   \institute{Donostia International Physics Center, Manuel Lardizabal Ibilbidea, 4, 20018 Donostia, Gipuzkoa, Spain
        \and
             Euskal Herriko Unibertsitatea, Edificio Ignacio Maria Barriola, Plaza Elhuyar, 1, 20018 Donostia-San Sebastián, Spain
        \and Department of Astrophysics, University of Vienna, T\"urkenschanzstraße 17, 1180 Vienna, Austria
        \and
             IKERBASQUE, Basque Foundation for Science, 48013, Bilbao, Spain
             }

 
  \abstract{
  The intrinsic alignment of galaxies can provide valuable information for cosmological and astrophysical studies and is crucial for interpreting weak-lensing observations. Modelling intrinsic alignments requires understanding how galaxies acquire their shapes in relationship to the large-scale gravitational field, which is typically encoded in the value of large-scale shape-bias parameters. In this article, we contribute to this topic in three ways: (i) developing new estimators of Lagrangian shape biases, (ii) applying them to measure the shape biases of dark-matter halos, (iii) interpreting these measurements to gain insight into the process of halo-shape formation. Our estimators yield measurements consistent with previous literature and offer advantages over earlier methods, such as the independence of the bias measurements from other bias parameters and the ability to define bias parameters for each individual object. We measure universal relations between shape-bias parameters and peak height, $\nu$. For the first-order shape-bias parameter, this relation is close to linear at high $\nu$ and approaches zero at low $\nu$, which provides evidence against the proposed scenario that galaxy shapes arise due to post-formation interaction with the large-scale tidal field. We anticipate that our estimators will be very useful for analyzing hydrodynamical simulations to enhance our understanding of galaxy shape formation and for establishing priors on the values of intrinsic alignment biases.
  }

   \maketitle
%

\section{Introduction}

Cosmic shear (CS) is arguably the most direct way to probe the fluctuations of the matter field on cosmological scales \citep{Mandelbaum_2018}. The slight deformations to the baseline shapes of galaxies are not affected by the clustering properties of those galaxies, making these measurements independent of galaxy bias. As a result, the information that can be extracted on the amplitude of matter fluctuations, $\sigma_8$, does not have the typical degeneracy with linear galaxy bias. However, an outstanding issue is that the shapes of galaxies are not entirely random before shearing but may be correlated with the large-scale gravitational field. Typically known as galaxy intrinsic alignments (IA), this effect is highly dependent on galaxy-formation physics and contaminates cosmic-shear measurements \citep{Joachimi_2015, Kiessling_2015, Kirk_2015, Troxel_2015}. Therefore, extracting reliable information from CS requires accurate modelling of IA \citep{Secco_2022, Arico_2023, preston2024reconstructingmatterpowerspectrum}.

IA modelling is often based on a perturbative expansion \citep{Desjacques_2018}. Early works modelled galaxy shapes as being linearly correlated to the large-scale tidal field, with the linear-alignment parameter $c_K$ remaining free to describe any galaxy population \citep{Catelan_2001, Hirata_Seljak_2004}. In recent years, this field has seen many advancements, with the formulation of models including higher orders in perturbation theory (PT) \citep{Blazek_2015}, complete formulations using the effective field theory of large-scale structures (EFTofLSS) both in Eulerian \citep{Vlah_2020, Vlah_2021, Bakx_2023} and Lagrangian \citep{Chen_2024} space, and hybrid prescriptions which use a bias expansion combined with the results of $N$-body simulations to produce models that extend accurately into the non-linear regime \citep{Maion_2024}. An integral part of all of these models is the ``shape bias'' parameters -- free coefficients that encode the average response of galaxies' shape to large-scale properties of the matter field.

Although the bias parameters of a certain population of objects cannot be easily predicted \textit{a priori}, measuring their values can shed light on the formation mechanism of said objects. As an example, the detection of non-zero halo assembly bias for density bias parameters \citep{Gao_2005, Wechsler_2006, Gao_2007, Angulo_2008} showed that dependencies on halos' accretion histories had to be included in predictions from excursion-set theory \citep{BBKS, Bond_1991, Lacey_1993} to be in agreement with halo-bias measurements from $N$-body simulations \citep{Dalal_2008}. Furthermore, it explained why simple halo-occupation distribution (HOD) models, which assume the galaxy content of a halo to depend solely on its mass \citep{Zheng_2005}, are inadequate in describing realistic measurements from simulations and galaxy surveys \citep{Croton_2007, Zehavi_2018,Contreras:2024}. Measurements of shape biases are still in their infancy but should contribute similarly to the understanding of shape formation as they advance into a more mature stage.

A significant amount of research has been carried out to analyze intrinsic alignments of halos and galaxies, in both gravity-only and hydrodynamical simulations \citep{Tenneti_2014, Tenneti_2015, Tenneti_2015B, Tenneti_2016, Tenneti_2020, Chisari_2015, Velliscig_2015, Hilbert_2017, Delgado_2023}, but with little focus on large-scale shape-bias parameters. The first direct measurements of shape bias parameters of simulated halos have been performed by \cite{Stucker_2021} and \cite{Akitsu_2021} employing the ``anisotropic separate universe'' technique \citep{Schmidt_2018, Wagner_2014}. In a later work, \cite{Akitsu_2023} employed the quadratic bias technique \citep{Schmittfull_2015} to measure second-order shape biases. These works have shown that there is a power-law scaling of the linear shape bias with halo mass and demonstrated its dependence on concentration for high-mass halos. A universal relation between the linear shape bias and $b_1$, the linear density bias, was also established. Furthermore, they have quantified the dependence of second-order shape-bias parameters with halo mass and studied their dependence with $b_1$. Nevertheless, many interesting questions remain unanswered.

The precise origin of halo-shape alignments with the large-scale gravitational field is still poorly understood. In one possible scenario, the large-scale gravitational forces progressively deform halos after their formation, producing an alignment which grows over time. Another possibility is that the shape is already encoded during the formation of halos -- e.g. through the anisotropy of infall-velocities as determined by the large-scale gravitational potential. It should be possible to distinguish between these scenarios through measurements of the bias parameters. Beyond this, it is important to understand how alignments depend quantitatively on the (often arbitrary) boundary definition of the considered system and whether the alignment strength may be affected by secondary properties beyond halo mass -- such as the spin or formation time. 

While studies of halos in gravity-only simulations may provide a qualitative understanding of necessary model ingredients, the ultimate goal of theoretical alignment studies is an understanding of the shape biases of galaxies. Determining the values of those in realistic scenarios, such as high-resolution hydrodynamical simulations would be a great advancement towards obtaining informative priors that could be applied to IA models in weak-lensing analyses \citep{Zennaro_2022, ivanov2024fullshapeanalysissimulationbasedpriors, Ivanov_2024, Shiferaw_2024, zhang2024hodinformedprioreftbasedfullshape}. Additionally, the comparison of measurements of galaxy shape bias parameters obtained from hydrodynamical simulations to those measured from observations could be used as a way to constrain ``subgrid'' models of galaxy formation. To answer all these questions, we need a fast and precise estimator of shape biases.

In this manuscript, we generalize the theoretical formalism presented in \cite{Stucker_2024} to develop estimators of shape biases. The estimators are based on measurements of correlations in the joint probability distribution of Lagrangian large-scale tides and the shape of a given object. Similar to the separate universe approach, this yields simple and robust measurements of these biases but with the additional benefit of low computational cost, making the method viable for use in large hydrodynamical simulations. We validate the estimators through convergence tests and a comparison with literature results. Finally, we provide detailed measurements of the shape biases of halos and discuss the theoretical implications for the modelling of intrinsic alignments.

The article is structured as follows. In Section \ref{sec:probabilistic_bias}, we derive the new estimators based on considerations from probability theory. In Section \ref{sec:simulations}, we present the set of simulations we use to validate our results. In Section \ref{sec:results}, we test the robustness of our shape bias measurements, and we provide accurate fitting formulae. Finally, in Section \ref{sec:conclusions}, we summarize our findings and discuss future applications.

\section{Probabilistic Bias}
\label{sec:probabilistic_bias}

\begin{figure*}
    \centering
    \includegraphics[width=\textwidth]{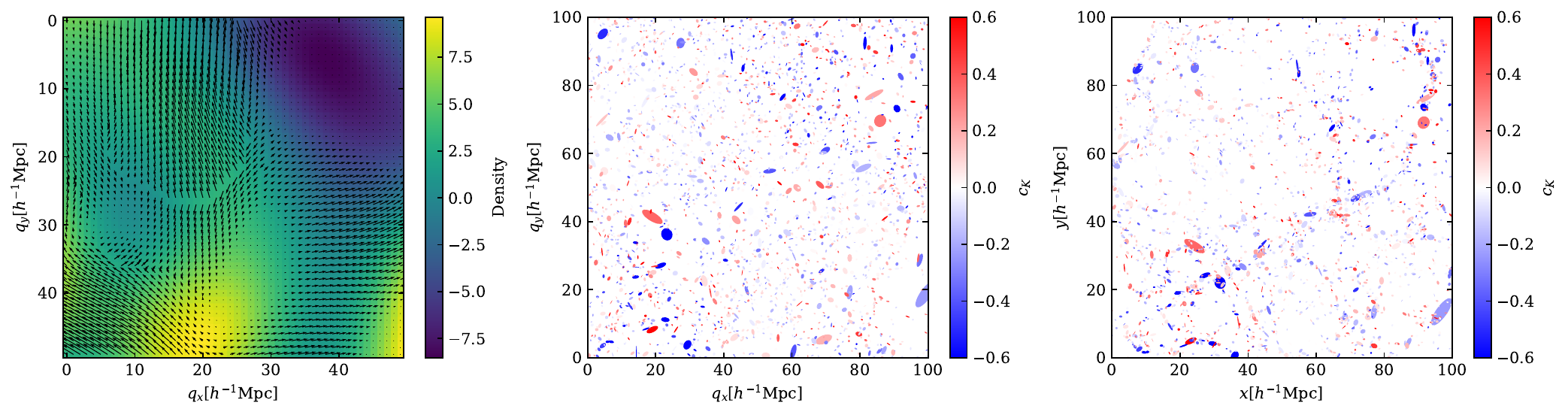}
    \caption{Illustration of halo shapes in a simulation slice and their relation to the underlying density and tidal fields. Panel one shows the linearly extrapolated density field to $z=0$ on the background, and superimposed, one can see a large set of small arrows; these small arrows have their size scaled to the local projected tidal field's largest eigenvalue, and their direction is that of the corresponding eigenvector. The second panel shows the halos represented as coloured ellipses. These ellipses are centred at the halos' Lagrangian positions, and their shapes are those of the projected shape tensor estimated for the FoF group; their colour corresponds to their value of the three-dimensional shape-alignment bias. Finally, the third panel is identical to the second, except that now the ellipses are placed at their final Eulerian positions. }
    \label{fig:enter-label}
\end{figure*}

Considering an object composed of a finite set of particles, $N$, one can define its shape tensor $\mathcal{I}$. This quantity is a rank-2 tensor, generally defined as

\begin{equation}
    \mathcal{I}_{ij} = \frac{1}{N}\sum_{n=1}^N \left(x_i^{(n)} - \overline{x}_i\right)\left(x_j^{(n)} - \overline{x}_j\right),
    \label{eq:shape_tensor}
\end{equation}

\noindent with the corresponding reduced shape tensor:

\begin{equation}
    \mathcal{\tilde{I}}_{ij} = \frac{1}{N}\sum_{n=1}^N \frac{\left(x_i^{(n)} - \overline{x}_i\right)\left(x_j^{(n)} - \overline{x}_j\right)}{r_n^2},
    \label{eq:red_shape_tensor}
\end{equation}

\noindent where $x_i^{(n)}$ represents the Eulerian position of each particle,  $n$ is an index running over the particles in the set, and $i$ runs over the three Cartesian coordinates, $\overline{x}_i$ is the mean of $x_i^{(n)}$, and $r_n^2 \equiv \sum_i (x^{(n)}_i - \overline{x}_i)^2$. We will use the non-reduced shape definition for all results in this article, except when comparing different definitions.

Now, we can define the normalized and trace-free part of $\mathcal{I}$:

\begin{equation}
    I_{ij}(\mathbf{x}) = \frac{\mathcal{I}_{ij}(\mathbf{x})}{\mathcal{I}_0}  - \frac{1}{3}\delta^K_{ij},
\end{equation}

\noindent in which $\mathcal{I}_0$ is the trace of $\mathcal{I}_{ij}$, and $\delta^K_{ij}$ is the Kronecker delta symbol. 

The tensor $I$ describes the shape of an object, factoring out its physical size. Being a symmetric tensor, it can be diagonalized to give three eigenvalues $I_{1,2,3}$, such that $\sum_{i=1}^3 I_i=0$. Each of these measures the relative extension of the object in each direction of an orthogonal basis formed by its eigenvectors. The goal of this work is to understand the response of this shape to changes in the large-scale gravitational field.

Attempting to determine which large-scale properties could generate an effect, we find that the simplest object we can build, allowed by the equivalence principle and symmetry of the problem, is the Lagrangian tidal field:

\begin{equation}
    T_{ij}(\mathbf{q}) =\partial_i\partial_j \Phi_G(\mathbf{q}),
    \label{eq:tidal_field}
\end{equation}

\noindent where $\Phi_G$ is the displacement potential, $\mathbf{q}$ is a position in Lagrangian coordinates, and the derivatives are taken with respect to this coordinate system. For convenience, we also define the traceless part of the tidal field

\begin{equation}
    K_{ij}(\mathbf{q}) = \left( \partial_i\partial_j - \frac{1}{3}\delta_{ij}^K \right)\Phi_G.
\end{equation}

Assuming for now that the tidal field is the only property of the Lagrangian matter field affecting the shapes of gravitationally bound objects, we can now develop a theoretical framework to quantify its effect. 

A key assumption of the forthcoming calculations will be the validity of the peak-background split (PBS) \citep{BBKS}. That is, we will assume that the criterion for forming a halo will be defined relative to small-scale quantities; large scales will only affect this process by introducing modulations to the distribution of small-scale perturbations. Separate-universe (SU) simulations provide an exact implementation of this scenario \citep{Gnedin_2011, Wagner_2014, Schmidt_2018, Stucker_2021, Akitsu_2021}, and it is interesting to consider them for a moment. 

Consider a simulation where a constant perturbation $\mathbf{T}_0$ to the tidal field is introduced. Upon identifying a set of objects (e.g. halos or galaxies) in these simulations, one can compute their average shape as a function of the large-scale tide $\mathbf{T}_0$. This yields a functional relation $\left\langle\mathbf{I}|\mathbf{T}_0\right\rangle_g$ -- where our notation indicates an expectation value that has been conditioned on $\mathbf{T}_0$ and that has been taken over galaxies (or halos). This function can be expanded in a perturbative series
\begin{align}
    \left\langle\mathbf{I}|\mathbf{T}_0\right\rangle_g & = \mathbf{C}_{K,1} \mathbf{T}_0 + \mathbf{T}_0^T \mathbf{C}_{K,2}\mathbf{T}_0+\cdots,
    \label{eq:true_bias_function}
\end{align}

\noindent where $\mathbf{C}_{K,1}$ is a rank 2 tensor, $\mathbf{C}_{K,2}$ is of rank 4 and each product corresponds to a contraction over two indices. Therefore, by running SU simulations at several values of $\mathbf{T}_0$, one can compute derivatives of $\langle \mathbf{I}|\mathbf{T}_0\rangle$, and hence extract generalized tensorial bias parameters:

\begin{equation}
    \mathbf{C}_{K,n} = \frac{\partial^n\langle\mathbf{I}|\mathbf{T}_0\rangle}{\partial \mathbf{T}_0^n}\Big|_{\mathbf{T}_0=0}.
    \label{eq:PBS_bias_def}
\end{equation}

SU simulations, therefore, provide a well-defined route to compute shape bias parameters. Unfortunately, they have the downside of being costly due to the need to run several simulations at different values of $\mathbf{T}_0$. We will now show an alternative route to measure these bias parameters, fully consistent with the SU method, using quantities that can be obtained from only a single simulation.

We begin by considering the linear field of a simulation in a finite volume, smoothed with a sharp filter in $k$-space such that only modes with $k<k_d$ are retained. This linear field is characterized at first order by its tidal field, which we denote as $\mathbf{T}$. 

If we average the shape of galaxies that inhabit regions of the same tidal field $\mathbf{T}$, we obtain the conditional expectation value $\langle \mathbf{I}|\mathbf{T} \rangle_g$. Note that this function is slightly different than $\langle \mathbf{I}|\mathbf{T}_0 \rangle_g$, since the tidal field measured at the finite scale $k_d$ carries a different amount of information than the idealized infinite scale field $T_0$. However, using probability theory, we can emulate a change in the field at infinite scales, which modifies the distribution of the field at finite scales.

First of all, consider that if we have an expectation value of some variable $h$ that is conditioned on two probability variables $\langle h | x,y \rangle$, then we can obtain the expectation value that is only conditioned on the variable $x$ through a weighted average as

\begin{align}
    \langle h | x\rangle &= \frac{\int \langle h | x,y \rangle p(y | x) \mathrm{d}y}{\int p(y|x) \mathrm{d}y}
\end{align}

\noindent where the denominator can be omitted if $p(y|x)$ is a normalized distribution. Additionally, we need to consider that the distribution of interest is not the (Gaussian) conditional distribution at random locations -- let us refer to it as $p(\mathbf{T} | \mathbf{T}_0)$ -- but instead it is the distribution at the locations of galaxies $p_g(\mathbf{T} | \mathbf{T}_0)$. The two are related through a bias function:

\begin{align}
    p_g(\mathbf{T} | \mathbf{T}_0) &= f(\mathbf{T}, \mathbf{T}_0) p(\mathbf{T} | \mathbf{T}_0),
\end{align}

\noindent where $f(\mathbf{T}, \mathbf{T}_0)$ describes the relative excess probability of finding a galaxy in a region that has the tidal field $\mathbf{T}$ at finite scales and $\mathbf{T}_0$ at infinite scales \citep[see][for a detailed explanation]{Stucker_2024}:
\begin{equation}
    f(\mathbf{T},\mathbf{T}_0) = \frac{p(g|\mathbf{T},\mathbf{T}_0)}{p(g)},
    \label{eq:dens_bias_function}
\end{equation}

or, through Bayes' theorem, 

\begin{equation}
    f(\mathbf{T}, \mathbf{T}_0) = \frac{p(\mathbf{T},\mathbf{T}_0|g)}{p(\mathbf{T},\mathbf{T}_0)}.
\end{equation}

Under the PBS assumption the value of $\mathbf{T}_0$ is irrelevant if the smaller scale field $\mathbf{T}$ is known, so that $f(\mathbf{T}, \mathbf{T}_0) = f(\mathbf{T})$. Putting things together, we find:

\begin{align}
    \langle \mathbf{I} | \mathbf{T}_0 \rangle_g &= \frac{\int \langle \mathbf{I} | \mathbf{T}_0, \mathbf{T} \rangle_g f(\mathbf{T}) p(\mathbf{T} | \mathbf{T}_0) d \mathbf{T} }{ \int f(\mathbf{T}) p(\mathbf{T} | \mathbf{T}_0) d \mathbf{T} }
\end{align}

We may further simplify this expression by using that, under the PBS assumption, $\langle \mathbf{I} | \mathbf{T}_0, \mathbf{T} \rangle_g = \langle \mathbf{I} | \mathbf{T} \rangle_g$. Further, we define the ``renormalized bias function'':

\begin{align}
    F(\mathbf{T}_0) &= \int f(\mathbf{T}) p(\mathbf{T} | \mathbf{T}_0) d \mathbf{T} \nonumber \\
       &=  \langle f(\mathbf{T} + \mathbf{T}_0) \rangle
\end{align}

\noindent where the second line assumes the use of a sharp k-space filter to define $\mathbf{T}$ \citep{Stucker_2024}. We then find

\begin{align}
\label{eq:small_to_large}
    \langle \mathbf{I} | \mathbf{T}_0 \rangle_g &= \frac{1}{F(\mathbf{T}_0)} \int \langle \mathbf{I} | \mathbf{T} \rangle_g f(\mathbf{T}) p(\mathbf{T} | \mathbf{T}_0) d \mathbf{T} \\
    &= \frac{1}{F(\mathbf{T}_0)} \int \int \mathbf{I} \frac{p(\mathbf{I}, \mathbf{T} | g)}{p(\mathbf{T} | g)}  f(\mathbf{T}) p(\mathbf{T} | \mathbf{T}_0) d\mathbf{I} d \mathbf{T} \\
    &= \frac{1}{F(\mathbf{T}_0)} \int \int \mathbf{I} p(\mathbf{I}, \mathbf{T} | g) \frac{p(\mathbf{T} | \mathbf{T}_0)}{p(\mathbf{T})} d\mathbf{I} d \mathbf{T} \\
    &= \frac{1}{F(\mathbf{T}_0)} \left\langle \mathbf{I} \frac{p(\mathbf{T} | \mathbf{T}_0)}{p(\mathbf{T})} \right\rangle_g,
    \label{eq:small_to_large_final}
\end{align}

\noindent where in the second line we expanded $\langle \mathbf{I}|\mathbf{T} \rangle_g$ as an integral, in the third line we used Bayes' theorem $f(\mathbf{T}) = p(g|\mathbf{T})/p(g) = p(\mathbf{T}|g)/p(\mathbf{T})$, and we rewrote the double integral over $\mathbf{I}$ and $\mathbf{T}$ weighted by $p(\mathbf{I},\mathbf{T}|g)$ as an average over all galaxies. Equation (\ref{eq:small_to_large_final}) simply tells us that one can recover the SU average $\langle\mathbf{I}|\mathbf{T}_0\rangle_g$ by calculating the weighted average of shapes -- where the weight corresponds to the relative increase in the probability of finding each galaxy's environment if a large-scale tidal field $\mathbf{T}_0$ is applied.

For the case of a sharp $k-$space filter, it is $p(\mathbf{T} | \mathbf{T}_0) = p(\mathbf{T} - \mathbf{T}_0)$ so that after substituting into \eqref{eq:PBS_bias_def} we find:

\begin{align}
    \label{eq:c_k1_bias_third}
    \mathbf{C}_{K,1} & = \left\langle \frac{1}{p(\mathbf{T})}\mathbf{I}\otimes \left( -\mathbf{p}^{(1)}(\mathbf{T}) -  p\mathbf{F}^{(1)}(0) \right)  \right\rangle_g\\
    \label{eq:C_k2}
    \mathbf{C}_{K,2} & = \left\langle \frac{1}{p(\mathbf{T})}\mathbf{I}\otimes \left( -\mathbf{p}^{(2)}(\mathbf{T}) + 2\mathbf{p}^{(1)} \otimes \mathbf{F}^{(1)} - p\mathbf{F}^{(2)} + p\mathbf{F}^{(1)} \right)  \right\rangle_g,
\end{align}

\noindent where we write the derivatives with respect to $\mathbf{T}$ in boldface to remark that this derivative is a $2k$-rank tensor in which $k$ is the order of the derivative. 

The procedure outlined in this section defines an estimator of the shape biases, which we will show to be unbiased and which has some notable properties:

\begin{itemize}
    \item It can be defined on an object-by-object basis, by evaluating equation (\ref{eq:c_k1_bias_third}) without the averaging brackets;
    \item This estimator does not rely on the calculation of $n$-point functions, making it simpler to evaluate for any selection of objects, even in complex geometries;
    \item The renormalization procedure is conceptually very simple in this picture.
\end{itemize}
These properties will also be present in all of the estimators to be derived in the following sections of this manuscript.

\subsection{Linear Bias}
\label{sec:linear_bias}

The derivation leading to equation (\ref{eq:c_k1_bias_third}) has been left so far in a very general form; we will now connect it to more common definitions of shape biases found in the literature.

From equation (\ref{eq:PBS_bias_def}) taken at $n=1$, we can see that the values assumed by the rank-4 tensor $\mathbf{C}_{K,1}$ are constrained by the symmetries of the RHS, that is, $\mathbf{C}_{K,1}$ has to be isotropic, symmetric over its first two indices (given the symmetry of $\langle \mathbf{I}|\mathbf{T}_0\rangle$) and the last-two as well (given the symmetry of $\mathbf{T}$). In Appendix \ref{sec:appendix_IT} we summarize the findings of \cite{Stucker_2024}, who derived an orthogonal basis for all isotropic rank-4 tensors which respect the imposed symmetries,  $\mathds{V}_{22}=\mathrm{span}(\{\mathbf{J}_{22}, \mathbf{J}_{2=2}\})$, in which the tensors $\mathbf{J}_{22}$ and $\mathbf{J}_{2=2}$ are defined by

\begin{align}
    \mathbf{J}_{22} &= S_{22}(\delta^K_{ij}\delta^K_{kl})\\
    \mathbf{J}_{2=2} &= S_{22}(\delta^K_{ik}\delta^K_{jl}) - \frac{1}{3}S_{22}(\delta^K_{ij}\delta^K_{kl}),
\end{align}

\noindent where $S_{22}$ is a double symmetrization operator, defined via its action on an arbitrary rank-4 tensor $\mathbf{M}$, given by
\begin{equation}
    S_{22}(M_{ijkl}) = \frac{1}{4}\left( M_{ijkl} + M_{jikl} + M_{ijlk} + M_{jilk} \right).
\end{equation}

It is useful to keep in mind that the notation we employ gives information on the result of contracting these tensors with any arbitrary rank-2 tensors $\mathbf{A}$ and $\mathbf{B}$. Contracting $\mathbf{J}_{22}$ with $\mathbf{A}$ and $\mathbf{B}$ will simply give us the product of their traces -- each subscript $``2"$ without any connection symbol represents taking the trace of a rank-2 tensor. As for $\mathbf{J}_{2=2}$, contracting it with $\mathbf{A}$ and $\mathbf{B}$ will give the trace of their matrix product; each $``-"$ symbol connecting the subscripts represents summation over one repeated index; in this case, the first one is due to the matrix product, and a second one must appear to form the trace of this product.

Using this basis, one can uniquely decompose $\mathbf{C}_{K,1}$ as a linear combination,

\begin{equation}
    \mathbf{C}_{K,1} = c_{J_{22}} \mathbf{J}_{22} + c_{J_{2=2}} \mathbf{J}_{2=2},
    \label{eq:first_order_bias}
\end{equation}

\noindent and, by contracting with the basis tensors and taking advantage of orthogonality, we can write the expressions for the bias parameters in equation \eqref{eq:first_order_bias},
\begin{align}
    \label{eq:bias_1st_order_def1}
    c_{J_{22}} & = -\left\langle \frac{1}{p(\mathbf{T})} \left(\frac{\partial p}{\partial \mathbf{T}} -p\frac{\partial F}{\partial \mathbf{T}}(0) \right) \otimes \mathbf{I} \overset{(4)}{\cdot}\frac{\mathbf{J}_{22}}{||\mathbf{J}_{22}||^2}  \right\rangle_{\mathrm{g}}\\
    c_{J_{2=2}} & = - \left\langle \frac{1}{p(\mathbf{T})}\left(\frac{\partial p}{\partial \mathbf{T}} - p\frac{\partial F}{\partial \mathbf{T}}(0) \right) \otimes \mathbf{I} \overset{(4)}{\cdot}\frac{\mathbf{J}_{2=2}}{||\mathbf{J}_{2=2}||^2}  \right\rangle_{\mathrm{g}}.
    \label{eq:bias_1st_order_def2}
\end{align}

To evaluate these expressions, we need to examine the form of the PDF for the Lagrangian tidal field $p(\mathbf{T})$. \cite{Stucker_2024} showed that the p.d.f. for the tidal field can be written as,
\begin{equation}
        p(\mathbf{T}) = N \exp\left[ -\frac{1}{2}\mathbf{T} \mathbf{C}_T^\dagger \mathbf{T} \right]
\end{equation}

\noindent where $N$ is a normalization and $\mathbf{C}_T^\dagger$ is the pseudo-inverse of the covariance of the tidal field, given by

\begin{equation}
    \mathbf{C}_T^\dagger = \frac{1}{\sigma^2}\mathbf{J}_{22} + \frac{15}{2\sigma^2}\mathbf{J}_{2=2}
\end{equation}

\noindent where $\sigma$ is the variance of the density-field, damped at the scale of choice. From this expression, one can straightforwardly compute the first derivative of $p$ with respect to the tidal field, given by

\begin{equation}
    \frac{1}{p}\frac{\partial p}{\partial \mathbf{T}} = -\mathbf{C}_T^\dagger\mathbf{T}.
    \label{eq:first_derivative}
\end{equation}

Therefore, one can rewrite equations (\ref{eq:bias_1st_order_def1}, \ref{eq:bias_1st_order_def2}) as 
\begin{align}
    \label{eq:derived_bias_1st_order_def1}
    c_{J_{22}} & = \left\langle \mathbf{C}_T^\dagger \mathbf{T} \otimes \mathbf{I} \overset{(4)}{\cdot}\frac{\mathbf{J}_{22}}{||\mathbf{J}_{22}||^2}  \right\rangle_{\mathrm{g}} = \frac{1}{\sigma^2}\left\langle\mathrm{Tr}(\mathbf{T})\mathrm{Tr}(\mathbf{I})\right\rangle_g = 0 \\
    c_{J_{2=2}} & = \left\langle \mathbf{C}_T^\dagger \mathbf{T} \otimes \mathbf{I} \overset{(4)}{\cdot}\frac{\mathbf{J}_{2=2}}{||\mathbf{J}_{2=2}||^2}  \right\rangle_{\mathrm{g}} = \frac{3}{2\sigma^2}\left\langle \mathrm{Tr}(\mathbf{K}\mathbf{I})\right\rangle_g
    \label{eq:derived_bias_1st_order_def2}
\end{align}

\noindent in which the first term is equal to zero because we defined $\mathbf{I}$ to be trace-free; the first bias parameter, $c_{J_{2=2}}$ is exactly equivalent to the more commonly defined $c_{K}$, that is, the response of galaxy shapes to perturbations in the traceless part of the tidal field. Notice that 

\subsection{Second-Order Biases}

We can see from equation (\ref{eq:true_bias_function}) that the rank-6 tensor $\mathbf{C}_{K,2}$ is constrained by the symmetries to be isotropic and symmetric over its first, second, and third pairs of indices. As explained in Appendix \ref{sec:appendix_IT}, the set of isotropic rank-6 tensors which respect the imposed symmetries can be generated by the following orthogonal basis
\begin{equation}
    \mathds{V}_{222} = \mathrm{span}(\left\{\mathbf{J}_{222}, \mathbf{J}_{2-2-2-}, \mathbf{J}_{2=22},
    \mathbf{J}_{22=2},
    \mathbf{J}_{222=} \right\}),
\end{equation}
implying that one can uniquely decompose $\mathbf{C}_{K,2}$ as a linear combination of these tensors
\begin{equation}
    \begin{split}    
        \mathbf{C}_{K,2} = c_{J_{222}} \mathbf{J}_{222} + c_{J_{22=2}} \mathbf{J}_{22=2} & + c_{J_{2=22}} \mathbf{J}_{2=22}\\
        &+ c_{J_{222=}} \mathbf{J}_{222=} + c_{J_{2-2-2-}} \mathbf{J}_{2-2-2-},
        \label{eq:first_order_bias}
    \end{split}
\end{equation}
and by contracting with the basis tensor one obtains the bias parameters
\begin{align}
    \label{eq:bias_2nd_order}
    c_{J_{2=22}} & = \left\langle \frac{1}{p(\mathbf{T})} \left( p \frac{\partial^2 F}{\partial \mathbf{T}^2} - 2 \frac{\partial p}{\partial \mathbf{T}} \frac{\partial F}{\partial \mathbf{T}} + \frac{\partial^2 p}{\partial \mathbf{T}^2} \right) \otimes \mathbf{I} \overset{(6)}{\cdot}\frac{\mathbf{J}_{2=22}}{||\mathbf{J}_{2=22}||^2}  \right\rangle_{\mathrm{g}},
\end{align}

\noindent where, for the rest of this work, we will assume that the 6-dimensional dot product is performed in such a way that $\mathbf{I}$ gets contracted with the last two indices of $\mathbf{J}_{X}$ and the first 4 are contracted with the derivatives of $p$ and $F$. This is relevant because terms, such as the one we have written, that have as their last index a disconnected $``2"$ subscript will be proportional to the trace of $\mathbf{I}$, which is zero. The same will not occur for terms such that the disconnected $``2"$ subscript will get contracted with the derivatives. We refrain from writing all of the terms for conciseness because they are completely analogous.

We now turn to the evaluation of the expression in parenthesis in Equation (\ref{eq:bias_2nd_order}). The second derivative of the p.d.f. of the tidal tensor can be obtained straightforwardly

\begin{align}
    \frac{1}{p}\frac{\partial^2 p}{\partial \mathbf{T}^2} & = \frac{1}{4}\left[ \mathbf{C}_T^\dagger(\mathbf{T}+\mathbf{T}^T) \right] \otimes \left[ \mathbf{C}_T^\dagger(\mathbf{T}+\mathbf{T}^T)\right] - \mathbf{C}_T^\dagger,
\label{eq:second_derivative}
\end{align}
and its first derivative has already been discussed in Equation (\ref{eq:first_derivative}). As for $F$, the density bias function, we can write it explicitly as:

\begin{equation}
    F(\mathbf{T}) = 1 + \mathbf{B}_{T,1}\mathbf{T} + \frac{1}{2}\mathbf{T}\mathbf{B}_{T,2}\mathbf{T} + \frac{1}{6}\mathbf{B}_{T,3}\mathbf{T}\mathbf{T}\mathbf{T} + \cdots,
    \label{eq:dens_bias_fun}
\end{equation}
in which $\mathbf{B}_{T,N}$ are the generalized $N$-th order density bias parameters, which are defined as 
\begin{align}
    \mathbf{B}_{T,1} & = b_1\mathbf{J}_2\\
    \mathbf{B}_{T,2} & = b_2\mathbf{J}_{22} + 2b_{K^2}\mathbf{J}_{2=2}\\
    \mathbf{B}_{T,3} & = b_3\mathbf{J}_{222} + b_{\delta K^2}\mathbf{J}_{22=2} + b_{K^3}\mathbf{J}_{2-2-2-},
\end{align}
such that its derivatives will be given by
\begin{align}
\label{eq:first_F_derivative}
\frac{\partial F}{\partial \mathbf{T}}(0) &= \mathbf{B}_{T,1}\\
\label{eq:second_F_derivative}
\frac{\partial^2 F}{\partial \mathbf{T}^2}(0) &= \mathbf{B}_{T,2}.
\end{align}

Not all of these tensors will provide relevant contributions when substituted into equation (\ref{eq:bias_2nd_order}). Notice that, according to the convention adopted in this work, the last 2 indices of these rank-6 tensors will be contracted with $\mathbf{I}$, and therefore, $\mathbf{J}_{222}$ and $\mathbf{J}_{2=22}$ are guaranteed to be zero since they would extract the trace of $\mathbf{I}$, which is null. Furthermore, $\mathbf{J}_{2=22}$ and $\mathbf{J}_{222=}$ will result in identical results due to the symmetry in permutation of the first and second pairs of indices in equation (\ref{eq:bias_2nd_order}). Therefore, the relevant tensors, which define our second-order bias parameters, are

\begin{equation}
{\mathbf{J}_{2=22}, \mathbf{J}_{2-2-2-}}.
\end{equation}

Substituting the results of equations (\ref{eq:first_derivative},\ref{eq:second_derivative},\ref{eq:first_F_derivative},\ref{eq:second_F_derivative}) into equation (\ref{eq:bias_2nd_order}) we can work out the expressions for each of these bias parameters. These calculations are quite cumbersome, therefore we have developed a semi-numerical method for computing them. We simply quote the results obtained through those calculations,

\begin{align}
    \label{eq:estimators_c_dK}
    c_{\delta K} & = \frac{3}{2}\left\langle \delta \frac{\mathrm{Tr}(\mathbf{K}\mathbf{I})}{\sigma^4} \right\rangle_g - b_1c_K\\
    c_{K\otimes K} & = \frac{135}{7}\left\langle \frac{\mathrm{Tr}(\mathbf{K}\mathbf{K}\mathbf{I})}{\sigma^4} \right\rangle_g.
    \label{eq:estimators_c_KK}
\end{align}
It is interesting to spend a moment on equation (\ref{eq:estimators_c_dK}). Notice that this estimator, unlike the two others, depends on quantities that are defined in an ensemble average, notably $c_K$ and $b_1$, implying that, with this definition, it is not possible to estimate this bias parameter on an object-to-object basis.

\subsection{Linear Density Bias}

We are also interested in measuring the linear density bias of halos, $b_1$, to understand its coevolution with the shape bias parameters. The method we used for measuring this parameter is very similar in spirit to the ones described in this article, and is discussed extensively in \citep{Stucker_2024}. We refer the reader to that article for more information and suffice with quoting the equation that defines the estimator for $b_1$,
\begin{equation}
    b_1 = \left\langle \frac{\delta}{\sigma^2}\right\rangle_g.
\end{equation}

\section{Simulations}
\label{sec:simulations}

The simulations employed in this work are gravity-only $N$-body simulations run using a modified version of \texttt{L-GADGET3} \citep{Springel_2005, Angulo_2021}, with parameters summarised in Table \ref{tab:sim_pars}. The larger simulation, which we will refer to as the ``Planck'' simulation, has a cosmology compatible with that found by \cite{Planck_2018}. As for the two smaller ones, they have cosmological parameters specifically designed to reduce the errors of emulators in cosmological parameter space \citep{Angulo_2021}. The initial conditions are created at $z=49$ using fixed amplitudes \citep{Angulo_2016}, and then evolved to $z=0$. At each output redshift, the code stores properties of Friends-of-Friends (FoF) and \verb|SUBFIND| groups, which are interpreted to be halos and subhalos, respectively. Properties of these halos and subhalos are then computed from these particle distributions, using many different definitions and criteria, which we make explicit in the following subsection.

\subsection{Shape Definitions}

\begin{table}
 \caption{Shape and mass definitions employed throughout this article. We will employ the non-reduced FoF shape definition for all the results except when comparing the effect of these shape definitions on the resulting shape-bias parameters \ref{fig:shape_comparison}.}
 \label{tab:halo_defs}
 \begin{tabular}{p{2cm}p{6cm}}
    \hline\hline
    Symbol & Definition \\
    \hline
    $I_{\rm FoF}$ & Shape computed from the particles belonging to the Friends-of-Friends groups. The FoF groups in this article will be composed of all particles that have at least one neighbour particle closer than a linking-length of $0.16$ times the mean inter-particle separation.\\
    \\
    $I_{\rm200,b}, M_{\rm200,b}$ & Once defined $r_{\rm200,b}$, the distance from the halo center at which the average enclosed density is 200 times the mean density of the Universe. We compute the halo shape or mass by considering all particles such that $r<r_{\rm200,b}$. The halo center is defined to be the minimum of the local gravitational potential. \\
    \\
    $I_{\rm200,c}$ & Same as above but for $r_{\rm200,c}$, defined as the radius at which the average enclosed density is 200 times the critical density of the Universe $\rho_c=3H^2/8\pi G$.\\
    \\
    $I_{\rm500,c}$ & Same as above but for $r_{\rm500,c}$, defined as the radius at which the average enclosed density is 500 times the critical density of the Universe.\\
    \hline
 \end{tabular}
\end{table}
Halos can be defined in many different ways, and the resulting measurements of quantities such as mass or shape will be highly dependent on the precise definition used. We list in Table (\ref{tab:halo_defs}) a few definitions of shapes which will be used throughout this manuscript. We will generally use the FoF definition for the halo shapes while using $M_{\rm200,b}$ as the mass of our halos. This choice was made for convenience, since these properties were natively computed in our simulations, while the other ones had to be recomputed in post-processing. 

\subsection{Shape Convergence}

Another important matter is to determine the minimum number of particles a halo should have so that its shape-measurement is well converged. We estimate this limit by measuring shape bias parameters for the two versions of the Narya simulation with different mass resolutions, described in Table \ref{tab:sim_pars}. The results of these measurements can be seen in Figure (\ref{fig:convergence_test}) of appendix \ref{sec:appendix_convergence}, and show that value of shape-biases are well-converged for halos with $M_h > 10^{12}M_{\odot}h^{-1}$. These halos are resolved with $\sim 270$ particles in the low-resolution Narya simulation. Consequently, in the remainder of this manuscript, we will only analyse halos resolved with at least 300 particles. 

\begin{table}
 \caption{Parameters employed in running the simulations used in this work. The parameters used for the \textit{Planck} simulation are compatible with those found by the Planck Collaboration \citep{Planck_2018}, and thus represent a standard $\Lambda$CDM cosmology. The choice of values for the parameters used in the Narya simulation is explained in \citep{Angulo_2021}. }
 \label{tab:sim_pars}
 \begin{tabular}{lccc}
    \hline\hline
    & Planck & Narya & Narya-HD \\ 
    \hline
    Parameter & Value & Value & Value \\
    \hline
    $h$ & $0.6774$ & $0.7$ & $0.7$\\
    $\sigma_8$ & $0.8159$ & $0.9001$ & $0.9001$ \\
    $\Omega_b$ & $0.0486$ & $0.05$ & $0.05$\\
    $\Omega_c$ & $0.2603$ & $0.31$ & $0.31$\\
    $n_s$ & $0.9667$ & $1.01$ & $1.01$\\
    $\tau$ & $0.0952$ & $0.0952$ & $0.0952$\\ 
    $N_{\mathrm{eff}}$ & $3.046$ & $3.046$ & $3.046$\\
    $L[$Mpc$/h]$ & $1024$ & $512$ & $512$\\
    $N_{\mathrm{part}}$ & $3072^3$ & $1536^3$ & $2288^3$\\
    \hline
 \end{tabular}
\end{table}

\subsection{Error Calculation}

Before entering into the description of our results, we describe how we estimate uncertainties in our measurements of the bias parameters. We adopt the Jacknife technique, dividing our simulation volume into $N_J = 4^3=64$ subregions, and then evaluate $N_J$ means of the bias parameters, each time leaving out one of the subvolumes. We compute the covariance and the mean of our estimate as
\begin{align}
    \mathbf{C} & = \frac{1}{N_J-1}\sum_n (\mathbf{b}_n - \mathbf{b}_0)\otimes(\mathbf{b}_n - \mathbf{b}_0)\\
    \mathbf{b}_0 & = \frac{1}{N_J}\sum_n \mathbf{b}_n,
\end{align}

\noindent where $\mathbf{b}_n$ represents a vector of the three bias parameters averaged in subregion $n$. Finally, the uncertainty on $c_K$ can be computed as $\sigma_{c_K} = \sqrt{\mathbf{C}_{00}}$. We will represent these uncertainties as errorbars that extend up and down from the points with a length of $\sigma_{c}$, except in Figure (\ref{fig:damping_scale_validation}) where we display these errors with a region filled one $\sigma_c$ upwards and one downwards of the average curve.

\section{Results} \label{sec:results}

In this section, we will present the results of applying our shape bias estimators to halos in the simulations described in the previous section. Before presenting the results, we emphasize that we can estimate $c_K$ and $c_{K\otimes K}$ for each individual object. This gives us great flexibility since once this set of biases is measured, we can average them over any quantity of interest, provided it is a small-scale property of the halos. To avoid confusion, we will state clearly in the figure caption the quantity used for binning, since it might not be the one being displayed. 

\subsection{Validation of Estimators}

\begin{figure*}
    \centering
    \includegraphics[width=\textwidth]{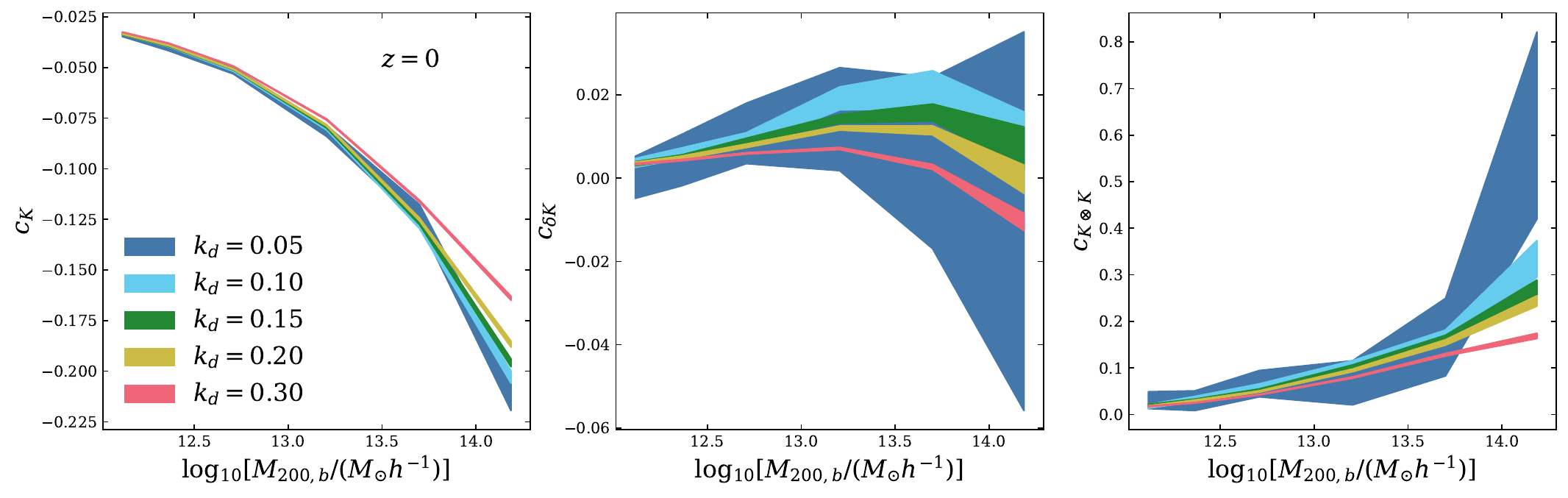}
    \caption{Halo bias parameters as a function of mass. Different colours indicate the damping scale $k_d$, measured in $[h$Mpc$^{-1}]$, used to smooth the Lagrangian fields employed in the measurement of the bias parameters, and the coloured regions indicate the 1-$\sigma$ interval around the mean value. Values presented here were measured for halos at $z=0$, and the measurements at higher redshift present even less dependence with $k_d$, such that any conclusion obtained here on the convergence of the estimators will also be valid at other times. }
    \label{fig:damping_scale_validation}
\end{figure*}

As described in Section \ref{sec:probabilistic_bias}, our bias estimators require smoothing the Lagrangian density field on an arbitrary scale $k_d$ in Fourier space. An important validation is to make sure that our estimates are, in fact, independent of $k_d$ if we make it sufficiently small (since bias parameters are formally scale-independent quantities). Figure (\ref{fig:damping_scale_validation}) shows the measurements of $c_K$, $c_{\delta K}$ and $c_{K\otimes K}$ averaged in bins of halo mass for increasing values of $k_d$. These measurements are displayed in the figure as shaded bands marking the 1-$\sigma$ region around the mean value of the biases. One can see that the measurements using $k_d=0.05[h$Mpc$^{-1}]$ and $k_d=0.1[h$Mpc$^{-1}]$ are compatible within 1-$\sigma$ over all masses, with the exception of the largest mass bin for $c_{K\otimes K}$, for which these estimators are compatible within 1.5-$\sigma$. Based on these checks, we will adopt $k_d=0.1[h$Mpc$^{-1}]$ to be the smallest scale at which our estimators are scale independent and use it to estimate all bias parameters that will appear in the following.

\begin{figure}
    \centering
    \includegraphics[width=\linewidth]{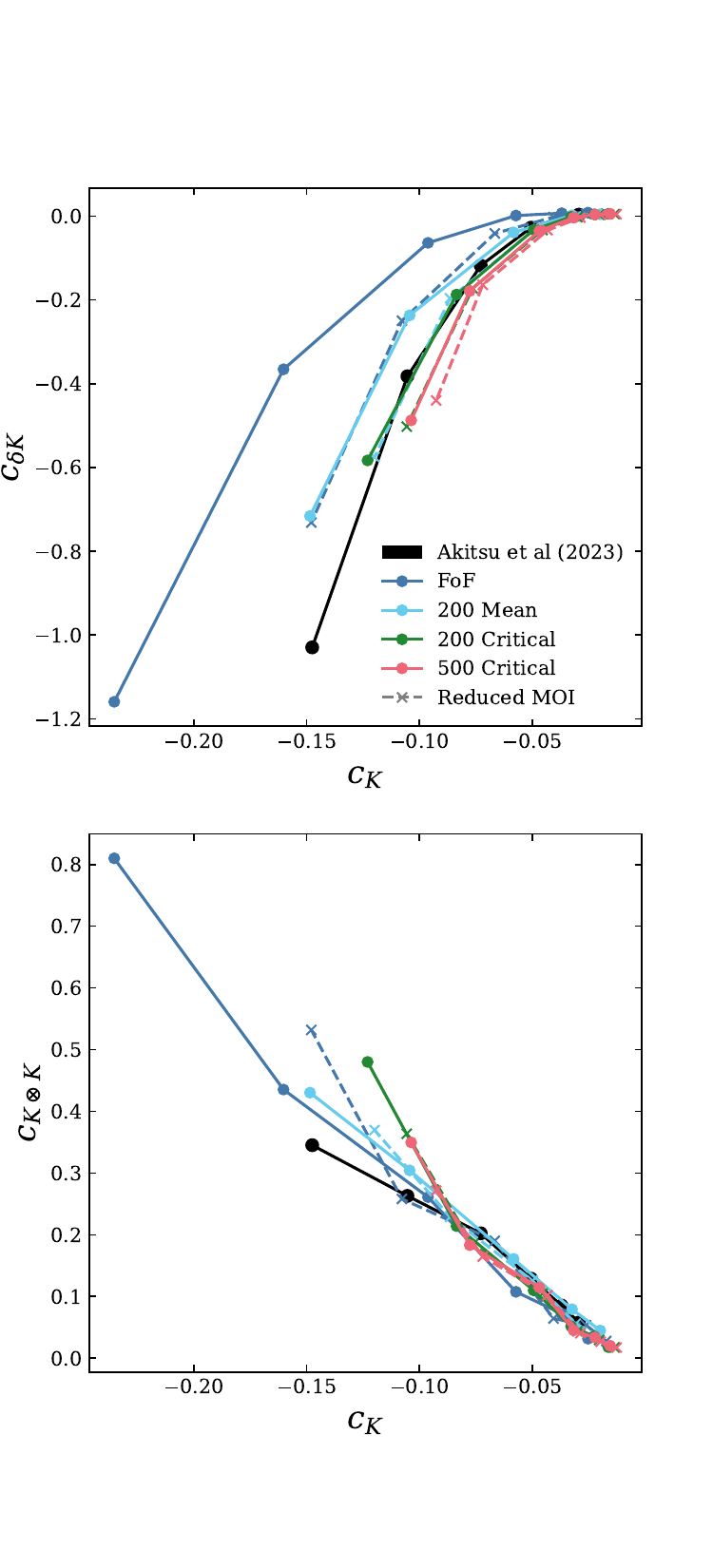}
    \caption{Comparison of shape-bias measurements performed with a range of different halo definitions to the measurements reported in \cite{Akitsu_2023}. There is a large variability in the relations found, and among them, $\tilde{\mathcal{I}}_{200,b}$ gives the best agreement with previous works.}
    \label{fig:shape_comparison}
\end{figure}

To validate our method, we compare the results to previous measurements in the literature \citep{Stucker_2021, Akitsu_2021, Akitsu_2023}. In Figure \ref{fig:shape_comparison}, we compare $c_{\delta K}$ and $c_{K\otimes K}$ versus $c_K$ for halos in different mass bins.  Different coloured lines represent different halo-shape definitions. Comparing the results for our standard definition, $I_{\rm FoF}$, to the measurements of \cite{Akitsu_2023} shows similar trends but a significant quantitative disagreement. However, the shapes of halos are expected to be very dependent on the halo definition used. Shape definitions based on spherical overdensity, such as $\mathcal{I}_{\rm 200,c}$, $\mathcal{I}_{\rm 500,c}$ and $\mathcal{I}_{\rm 200,b}$ will produce more isotropic halos, due to the fact that the particles belonging to the halo are pre-selected to be in a spherical region. On the other hand, the FoF shape definition will allow the halo outskirts to have varying shapes, which can heavily influence the measurement of $\mathcal{I}_{\rm FoF}$.

If we compare the measurements of \cite{Akitsu_2023} to our resukts for spherical boundary definitions -- in particular the reduced inertia tensor $\tilde{\mathcal{I}}_{\rm 200,b}$ --  we find a good agreement. Since these authors use the Amiga's Halo Finder (AHF) \citep{Knollmann_2009}, a spherical-overdensity based halo-finder, to define their halos and compute their bias parameters with the reduced shape tensor, this is the comparison that makes the most sense. The agreement is particularly good for $\tilde{I}_{200,b}$, which is unsurprising since those authors employ the same definition for the halo boundary. Therefore, we consider our measurements to be consistent with those reported by \cite{Akitsu_2023}, but we note that care must be taken when comparing different halo definitions -- much more so than e.g. when comparing halo density bias parameters. As the tidal alignment of halos is not directly observationally relevant, the preferential definition is not clear, and we stick to the FoF boundary for simplicity. However, we note that e.g. for measuring the alignment of galaxies in simulations it is advisable to use definitions that match observational criteria \citep{Tenneti_2015}.

\subsection{Universal Relations}

\begin{figure*}
\centering\includegraphics[width=\textwidth]{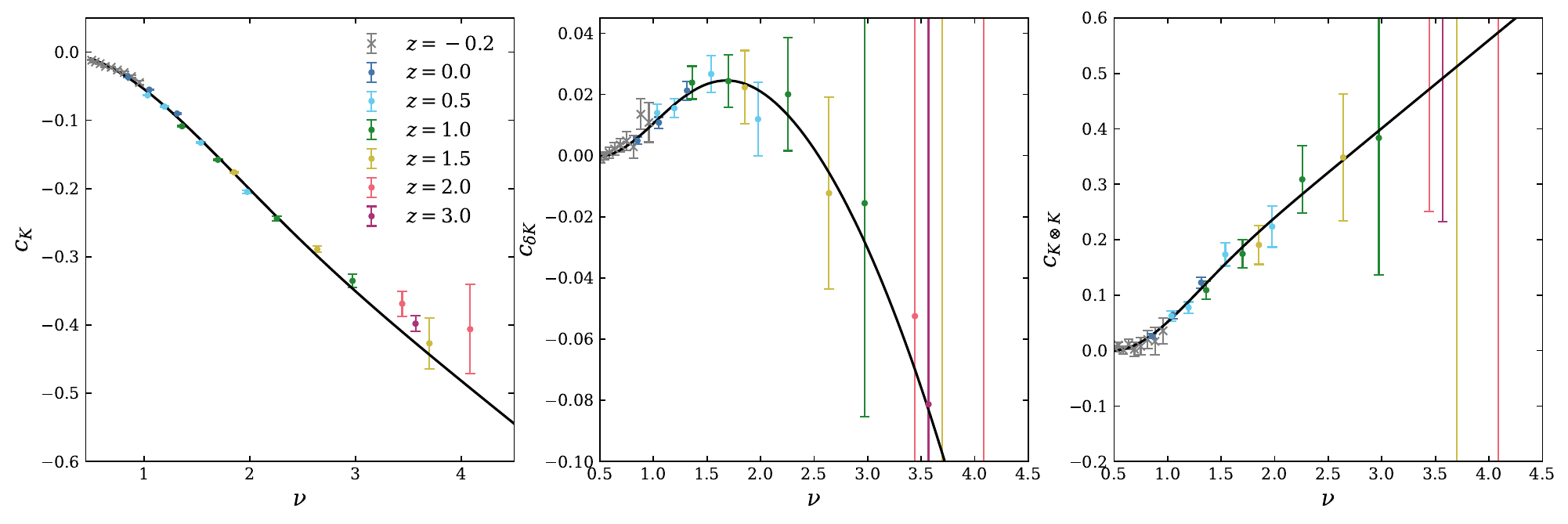}
    \caption{Measurements of shape bias parameters as a function of peak-significance $\nu=\delta_c/\sigma(M,z)$. All measurements, regardless of redshift, seem to follow a common trend with $\nu$. Black solid lines show the fits performed to approximate these common relations, and the smaller panel inside each figure shows the residues of the fit in units of the error in each bias estimate. These fitting functions are exceptionally simple -- linear for $c_K$ and $c_{K\otimes K}$ and quadratic for $c_{\delta K}$, and provide good descriptions to the measured relations.}
    \label{fig:biases_nu}
\end{figure*}

\begin{figure*}
    \centering
    \includegraphics[width=\textwidth]{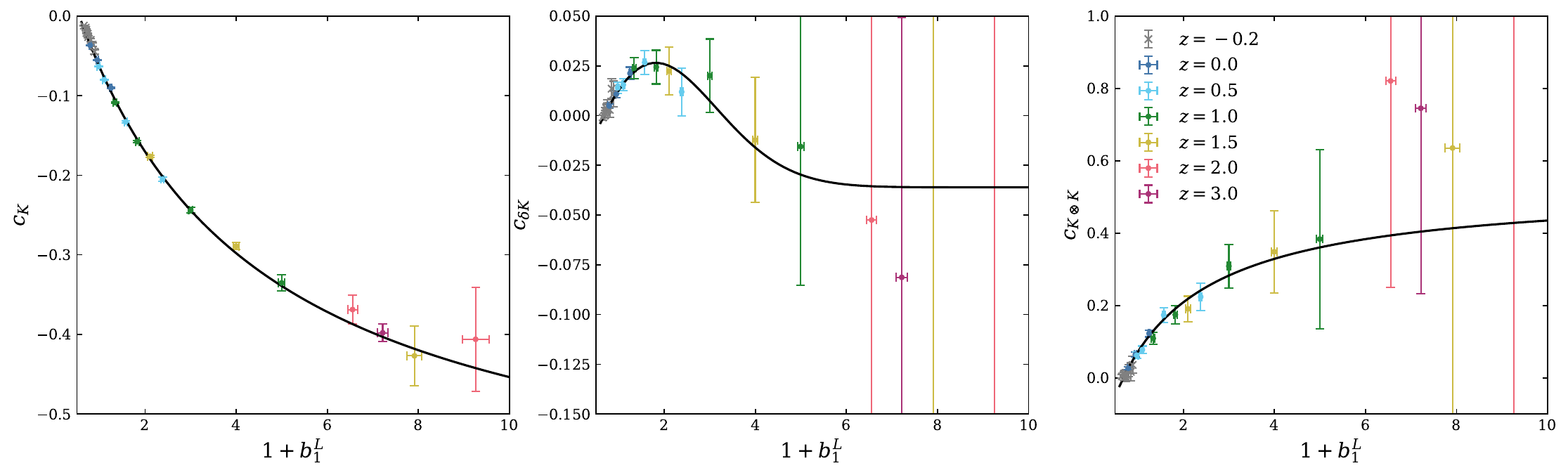}
    \caption{Shape biases averaged over bins in $\nu=\delta_c/\sigma(M,z)$, the peak-significance, as a function of linear density bias $b_1$ also averaged over bins of $\nu$. Different colors represent the redshifts indicated in the legend. Each large panel contains an inner panel that shows the orthogonal distance of the measurements to the fit, in units of the measurement error. The postulated functions provide a good description of the relations between the shape biases and $b_1$.}
    \label{fig:cX_per_b1}
\end{figure*}

As mentioned at the beginning of this section, our method has the advantage of being very flexible and allowing the measurement of shape bias parameters averaged over any quantity of choice. Figure (\ref{fig:biases_nu}) shows the shape bias parameters measured from the halos in the Planck and Narya-HD simulations, averaged directly in peak height $\nu=\delta_c/\sigma(M,z)$, where $\sigma(M,z)$ is defined as
\begin{equation}
    \sigma(M,z)^2 = \int \frac{d^3\mathbf{k}}{(2\pi)^3}P(k,z)W^2(kR),    
\end{equation}
where $R=\left( \frac{3M}{4\pi\bar{\rho}} \right)^{1/3}$, $\bar{\rho}$ is the mean density of the universe, and $W$ is a top-hat profile in Fourier space
\begin{equation}
    W(kR) = \frac{1}{(kR)^2}\left(\frac{3\sin(kR)}{kR} - \cos(kR)\right).
\end{equation}
This quantity is commonly used in excursion-set theory to measure how likely it is that the density in a region of radius $R$ has crossed $\delta_c$, the threshold for collapse. The relations in terms of $\nu$ are universal, that is, do not depend on redshift or cosmology. Furthermore, we notice that the functional forms relating the shape biases to $\nu$ can be described by simple linear or quadratic functions, which get damped at very low $\nu$, much simpler than those relating the shape biases to $b_1$. For comparison, we also show these relations with $b_1$ in Figure (\ref{fig:cX_per_b1}) with fits as listed in Table \ref{tab:biases_b1_fit_pars}.

\begin{table*}
 \caption{Compilation of functions used to fit the relations between shape biases and $\nu$, and the parameters of these functions which best fit the measurements. In the expressions for the functions being fitted, $\mathrm{erf}$ denotes the error function. }
 \label{tab:biases_nu_fit_pars}
 \centering
 \begin{tabular}{lccccc}
    \hline\hline
    & Function & a & b & c & d \\ 
    \hline
    $c_K$ & $\mathrm{erf}(dx - c) (ax + b)$ & $-0.12(7)$ & $0.0(2)$ & $0.1(7)$ & $0.57(9)$\\
    $c_{\delta K}$ & $\mathrm{erf}(x - d) (ax^2 + bx + c)$ & $-0.026(5)$ & $0.07(1)$ & $-0.03(1)$ & $0.56(6)$\\
    $c_{K \otimes K}$ & $\mathrm{erf}(dx - c)(ax + b)$ & $0.16(7)$ & $0.0(1)$ & $0.6(6)$ & $1.2(3)$\\
    \hline
 \end{tabular}
\end{table*}

The fact that $c_K$ depends strongly on $\nu$ is very revealing about the nature of alignments. Let us consider the hypothesis that halos get their alignments through the dynamical effect of the tidal field after formation. The tidal field may slightly distort the virialized halos -- shifting them to an anisotropic equilibrium configuration. We will refer to this as the ``post-formation-response'' scenario. Such a mechanism has been considered in several studies to explain the intrinsic alignment of galaxies \citep[e.g.][]{tugendhat_2018, zjupa_2020, ghosh_2024}, but it could also be a plausible effect for halos. 

If post-formation-response is the primary alignment mechanism of halos, then the degree of alignment should depend primarily on the anisotropy of the gravitational field in the vicinity of the halo. For a spherical halo with mass profile $M(r)$ plus a traceless large-scale tide $\mathbf{K}$, the gravitational acceleration is given by

\begin{align}
    \Vec{a}(\Vec{r}) &=  4 \pi G \rho_m \mathbf{K} \Vec{r} - \frac{G M(r)}{r^3} \Vec{r} \nonumber
\end{align}

\noindent where the factor $4 \pi G \rho_m$ is needed to correct for the dimensionless form of $\mathbf{K}$. If we consider an NFW halo at the virial radius $r_{200c}$, we may define the effective tidal tensor at the scale of the halo:
 
\begin{align}
    T_{\rm 200c} &= 4 \pi G \rho_{\rm m}  \mathbf{K} - \frac{G M_{\rm 200c}}{r_{\rm 200c}^3} \mathbf{J}_{2}  \nonumber \\
      &= 4 \pi G \rho_{\rm m} \left(\mathbf{K} - \frac{200}{3} \mathbf{J}_{2} \right),
\end{align}

\noindent where $\mathbf{J}_{2}$ is the unit matrix. The isotropic component is sourced purely by the halo, and the anisotropic component is purely by the large-scale tide. Therefore, the relative anisotropy in the gravitational field of an already formed halo is of the order $3 K_1 / 200$ (where $K_1$ is the largest eigenvalue of $\mathbf{K}$). The anisotropy in the triggered density response should be proportional to this and should probably also not be larger in amplitude. Therefore, we expect a post-formation-response that is mass-independent and of order $|c_K| \lesssim 10^{-2}$. This simple expectation is clearly in disagreement with the measured $c_K$ relations, which exhibit a strong mass (or $\nu$) dependence and a much larger degree of alignment up to $|c_K| \lesssim 0.5$. We conclude, therefore, that post-formation-response is not the dominant mechanism for the alignment of halos \citep[see][for a similar consideration for the case of galaxies]{camelio_2015}. The anisotropy must already be encoded during the formation and accretion processes.

We speculate about possible scenarios of how the $\nu$ dependence of $c_K$ may be encoded during formation. (1) A possible reason may be that the tidal field only significantly introduces anisotropy up to the formation of a halo and becomes rather irrelevant at later times. In this scenario, halos that form earlier have a smaller response because they have formed in a weaker tidal field. Combined with the fact that smaller $\nu$ leads to an earlier formation, this could qualitatively explain the observed trend. However, it is unclear whether this could match the measurements quantitatively and whether it is consistent with the observed universality. (2) Another possible reason for the observed $\nu$ dependence may be related to the fact that halos at low $\nu$ (and therefore large variance) are born in significantly larger tidal fields than halos at high $\nu$. The relative changes in the tidal tensor at the halo scale due to an imposed large-scale tidal field are, therefore, much smaller for small halos. (3) It might be that halo shapes are not responding as a continuous deformation to the large-scale tidal field, but are rather governed by discrete events -- e.g. through the formation of larger scale pancakes and filaments. The response of small halos may then be small because they are almost all embedded in larger-scale filaments, so it is unlikely that an increase in the tidal field triggers an additional filament formation. Such a process could be described through an excursion set formalism, which naturally leads to universality.

While so far a clear understanding of the formation of shapes is elusive, we note that there may be some hope to gain further insights: (a) The simple universal form of the response motivates the notion that a simple description may exist -- similar to the way that excursion sets offer a quite reasonable model for the density biases of halos. (b) All the above-mentioned scenarios could be turned into simple models with quantitative predictions that can be easily tested. (c) The response of halos could be investigated in warm dark matter simulations. The smallest warm dark matter halos have a clear and well-defined formation time and size \citep{Diemand_2005, Ishiyama_2010, Angulo_2017, Ogiya_2018, Ondaro-Mallea_2024}. It would be insightful to explore how such objects respond to tides before and after formation. (d) Some of the points above may be studied in anisotropic separate universe simulations. For example, for (2) to be correct, the differential response $|\partial/\partial T_0 \langle \mathbf{I} | \mathbf{T}_0 \rangle_g|$ at large values of the tidal field should be notably smaller than the response at $\mathbf{T}_0 = 0$.

\begin{table}
 \caption{Compilation of functions used to fit the relations between shape biases and $b_1$, and the parameters of these functions which best fit the measurements. }
 \label{tab:biases_b1_fit_pars}
 \begin{tabular}{lcccc}
    \hline\hline
    & Function & a & b & c \\ 
    \hline
    $c_K$ & $\frac{a + b x}{1+cx}$ & $0.056(4)$ & $-0.097(5)$ & $0.31(2)$\\
    $c_{\delta K}$ & $a + b x e^{cx^2}$ & $-0.018(4)$ & $0.029(5)$ & $-0.170(6)$\\
    $c_{K \otimes K}$ & $ax + b$ & $0.139(8)$ & $-0.076(9)$ & N/A\\
    \hline
 \end{tabular}
\end{table}

\subsection{Assembly Bias}

\begin{figure}
    \centering
    \includegraphics[width=\linewidth]{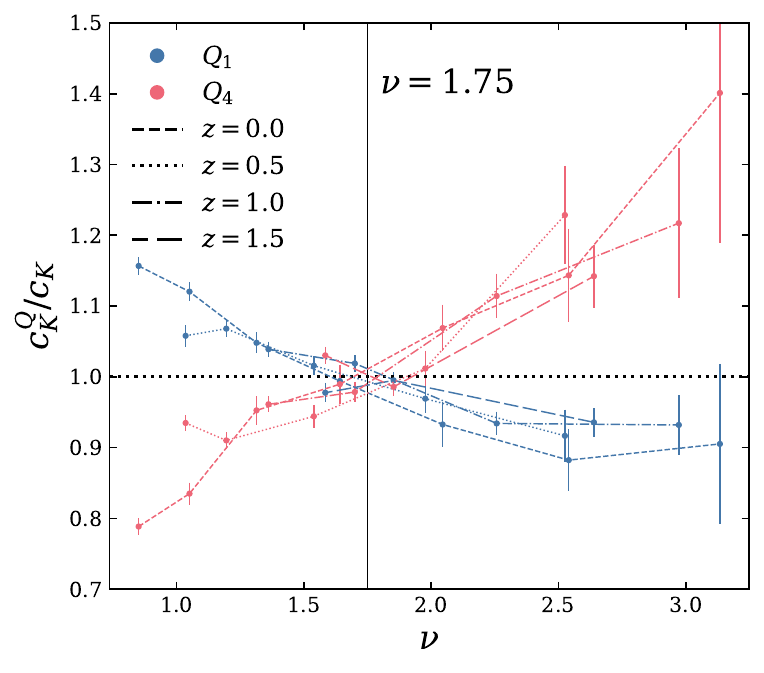}
    \caption{Measurements of linear shape bias, averaged over bins of $\nu$, for halos selected in 
    the lowest and highest quartiles in the distribution of spins. A clear assembly bias signal is detected such that, at low values of $\nu$, objects with low (high) spin have a $c_K$ value above (below) average; this behaviour is inverted around $\nu_*=1.75$, as marked by a vertical black line.}
    \label{fig:IA_bias_AB_spin}
\end{figure}

Besides the dependence of halo IA bias with $\nu$, it is interesting to explore whether if splitting the same $\nu$ populations by some secondary property, we find significantly distinct signals. We explore this possibility by splitting our samples according to their spin, which is defined as \citep{Bullock_2001}
\begin{equation}
    \lambda = \frac{|\mathbf{L}|}{\sqrt{2}v_{200,c}r_{200,c}},    
\end{equation}

\noindent where

\begin{equation}
    \mathbf{L} = \langle \mathbf{r} \times \mathbf{v} \rangle_{\mathrm{particles}},
\end{equation}

\noindent is the specific angular momentum vector of the object. We then split the population at each bin of $\nu$ into 4 quartiles of the spin distribution, such that $Q_1$ contains the halos with $25\%$ smallest spins and $Q_4$ the $25\%$ largest. Figure (\ref{fig:IA_bias_AB_spin}) shows the ratios between the biases of these sub-populations to the bias of the full sample. We can see that at low masses, the amplitude of $c_K^Q$ is anti-correlated with spin, such that lower-spin halos have higher bias and higher-spin halos have lower bias; this trend is inverted around $\nu\approx 1.75$, and beyond that the amplitude of $c_K$ is larger for high-spin halos and smaller for low-spin ones. As far as we are aware, this is the first time that a secondary dependence of $c_K$ on halo spin is observed.

\begin{figure}
    \centering
    \includegraphics[width=\linewidth]{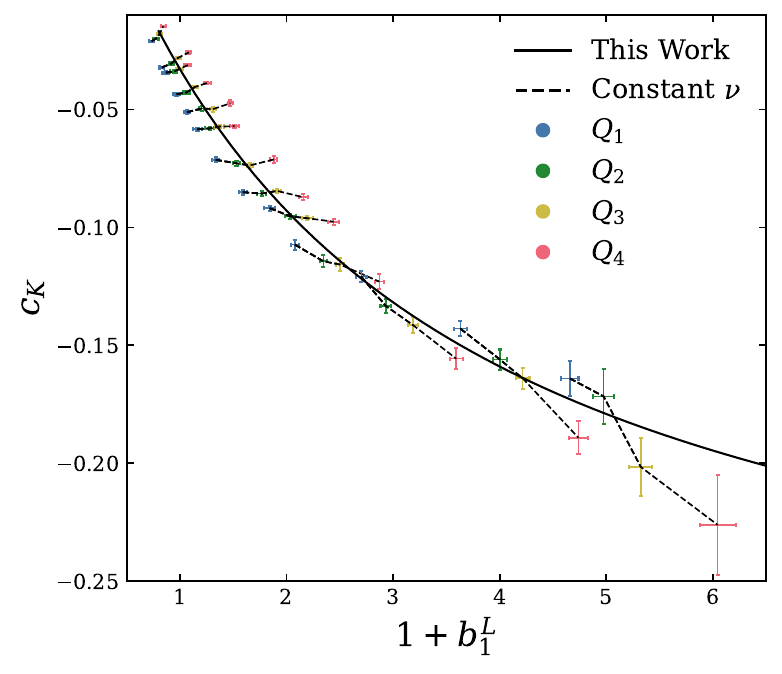}
    \caption{Linear shape bias as a function of linear density bias. Different colors represent the quartiles in the distribution of halo spins. One can clearly see a difference in the values of $c_K$ even at fixed $b_1$, demonstrating a non-trivial assembly-bias signal.}
    \label{fig:cK_b1_AB_nspin}
\end{figure}

It is well known that the linear density bias of halos $b_1$ has a secondary dependence on spin \citep{Gao_2007}, and hence it is interesting to probe whether the dependence of $c_K$ with spin is simply a consequence of the dependence of $b_1$ with spin. With that purpose, in Figure (\ref{fig:cK_b1_AB_nspin}) we show $c_K$ and $b_1$ averaged over $\nu$ bins and split into sub-populations by spin, as described earlier. One can clearly see the assembly bias in $b_1$ -- measurements that have the same $\nu$ are connected by dashed lines, and they have different values of $b_1$ depending on the spin-quartile they belong to. Nevertheless, one can also identify an independent $c_K$ assembly bias signal since $c_K$ does not vary along the $b_1 - c_K$ relation at fixed $\nu$ but varying spin. 

\section{Conclusions} \label{sec:conclusions}

In this article, we have presented a new methodology for computing shape biases in cosmological simulations from correlations in the distribution of the initial tidal tensor and the final shapes of objects. 

Our method is conceptually similar to the anisotropic separate-universe approach \citep{Stucker_2021, Akitsu_2021} and provides a robust measurement of the shape bias parameters that is independent of the number of considered parameters -- in notable contrast to methods that need to fit a model to power spectra or even field-level measurements \citep{Zennaro_2022, Schmittfull_2019}. We have shown that for sufficiently large damping scales, $k_d \leq 0.1h/$Mpc, our method is independent of the damping scale and matches the results from separate-universe simulations. However, the new method allows us to perform these measurements at a much-reduced computational cost for an arbitrary set of parameters from a single simulation -- making it viable for application in modern hydrodynamical simulations.

The application of these estimators to simulated halos has allowed us to quantify the dependence of the bias parameters $c_K$, $c_{\delta K}$ and $c_{K\otimes K}$ on peak-height $\nu$ and linear density bias $b_1$, and to provide fitting functions that accurately describe these relations. The strong mass dependence of the alignment response shows that the anisotropy must already be encoded during the formation and accretion history of objects. Post-formation-response -- as it is considered in some models of galaxy alignment \citep[e.g.][]{tugendhat_2018, zjupa_2020, ghosh_2024} -- is a too weak and too mass-independent effect to explain the linear large scale alignment of halos.

Further, we have presented the first detection of the dependence of the tidal alignment response $c_K$ on the spin of halos at a fixed peak height. This secondary dependence is rather complicated, leading to a larger response for higher spins at high masses but a smaller response for higher spins at lower masses. Understanding this relation may pose a significant challenge to analytical approaches to model alignments and tidal torquing \citep[e.g.][]{white_1984}.

The response of halo shapes to the large-scale tidal field provides important qualitative insights about collisionless systems and the ingredients that are necessary to model intrinsic alignments. However, quantitatively, it is far more relevant to understand the alignment of galaxies -- in particular, to correctly interpret weak-lensing observations. While we have validated our method in the case of halos in this article, it can easily be applied to simulated galaxies as well -- an avenue that we will be explore in future studies.

\begin{acknowledgements}
The authors thank Kazuyuki Akitsu for providing his measurements at our request. The authors acknowledge support from the Spanish Ministry of Science under grant number PID2021-128338NB-I00 from the European Research Executive Agency HORIZON-MSCA-2021-SE-01 Research and Innovation programme under the Marie Sklodowska-Curie grant agreement number 101086388 (LACEGAL). 
\end{acknowledgements}

\bibliographystyle{aa} 
\bibliography{prob_lag_shape} 

\begin{appendix}
    
\section{Notation}

In this section, we define some conventions in notation to be used throughout this work. This is particularly relevant because we will define some convenient ways to define operations between tensors that will appear commonly in this paper. Table \ref{tab:general_notation} resumes the definitions for most of the symbols employed throughout this article. Table \ref{tab:tensor_notation} resumes several of the most important notation for tensor operations used throughout this manuscript.

\begin{table}
 \caption{General notation employed throughout the article.}
 \label{tab:general_notation}
 \begin{tabular*}{\columnwidth}{@{}l@{\hspace*{50pt}}l@{\hspace*{50pt}}l@{}}
        \hline
        Notation & Definition\\
        \hline
        \\
        $\mathbf{x}$ & Eulerian positions\\
        $\mathbf{q}$ & Lagrangian positions\\
        $\mathbf{T}$ & Lagrangian tidal-field\\
        $\mathbf{K}$ & Traceless Lagrangian tidal-field\\
        $\mathbf{\mathcal{I}}$ & Shape tensor\\
        $\tilde{\mathbf{\mathcal{I}}}$ & Reduced shape tensor\\
        $\mathbf{I}$ & Traceless shape tensor\\
        $\tilde{\mathbf{I}}$ & Traceless reduced shape tensor\\
        $f$ & Finite-scale density bias function\\
        $F$ & Large-scale density bias function\\
        $p(\mathbf{T})$ & P.d.f of DM tidal field.\\
        $p_g(\mathbf{T})$ & P.d.f of DM tidal field at galaxy locations.\\
        $\mathbf{J}_2$ & $\delta K_{ij}^K$\\
        $\mathbf{J}_{22}$ & $S_{22}(\delta_{ij}\delta^K_{kl})$\\
        $\mathbf{J}_{2=2}$ & $S_{22}(\delta^K_{il}\delta^K_{kj})-\frac{1}{3}\mathbf{J}_{22}$\\
        $b_1$ & Lagrangian linear bias parameter.\\
        $c_K$ & Linear shape-bias parameter.\\
        $c_{\delta K},c_{K\otimes K}$ & Second-order shape-bias parameters.\\
        \\
        \hline
    \end{tabular*}
\end{table}

\begin{table}
 \caption{Notation for tensor operations, which will be employed throughout the calculations.}
 \label{tab:tensor_notation}
 \begin{tabular*}{\columnwidth}{@{}l@{\hspace*{50pt}}l@{\hspace*{50pt}}l@{}}
        \hline
        Compact Notation & Explicit Einstein Sum\\
        \hline
        \\
        $\mathbf{A}\mathbf{B}$ &  $A_{i_1i_2\alpha_1\cdots \alpha_n}B_{i_1i_2\beta_1\cdots \beta_m}$ \\
        \\
        $\mathbf{A}\overset{(N)}{\cdot}\mathbf{B} $ &  $A_{i_1i_2\cdots i_N\alpha_1\cdots \alpha_n}B_{i_1i_2\cdots i_N\beta_1\cdots \beta_m}$ \\
        \\
        $\mathbf{A}\otimes\mathbf{B}$ & $A_{i_1i_2\cdots i_n}B_{j_1j_2\cdots j_n}$\\
        \\
        $\frac{\partial \mathbf{A}}{\partial \mathbf{B}}$& $\frac{\partial A_{i_1i_2\cdots i_n}}{\partial B_{j_1j_2\cdots j_m}}$\\
        \\
        $||\mathbf{A}||^2$& $A_{i_1,i_2,\cdots,i_n}A_{i_1,i_2,\cdots,i_n}$\\
        \\
        \hline
    \end{tabular*}
\end{table}

\section{Isotropic Tensors}
\label{sec:appendix_IT}

In this appendix, we condense the findings of \citep{Stucker_2024} in finding an orthogonal basis for all isotropic tensors of rank $n$, that respect the following symmetry: let $M_{i_0i_1...i_{2n},i_{2n+1}...}$ be an isotropic tensor, then we require that $M$ is symmetric by the interchange $i_{2n}\leftrightarrow i_{2n+1}$. For tensors of rank $1$, there is only one such tensor, the Kronecker delta symbol
\begin{equation}
    \mathbf{J}_2 = \delta^K_{ij}.
\end{equation}
For rank 4 tensors, there are two such tensors,
\begin{align}
    \mathbf{J}_{22} & = \delta^K_{ij}\delta^K_{kl}\\
    \mathbf{J}_{2=2} & = S_{22}(\delta^K_{ij}\delta^K_{kl}) - \frac{1}{3}\mathbf{J}_{22},\\
\end{align}
in which the notation of the subscripts indicates properties of these tensors, namely that by contracting $\mathbf{J}_{22}$ with two rank-2 tensors $\mathbf{A}$ and $\mathbf{B}$, one will obtain the product of their traces -- so there will be no mixing of their components. By performing the same calculation with $\mathbf{J}_{2=2}$, on the other hand, one will obtain the trace of their matrix product. In fact, each symbol `-' in the subscripts represents one index contraction between tensors $\mathbf{A}$ and $\mathbf{B}$. Finally, for rank 6 there will be 5 different tensors
\begin{align}
    \mathbf{J}_{222} &= \delta^K_{ij}\delta^K_{kl}\delta^K_{mn}\\
    \mathbf{J}_{22=2} &= S_{222}(\delta^K_{ij}\delta^K_{kn}\delta^K_{ml}) -\frac{1}{3}\mathbf{J}_{222}\\
    \mathbf{J}_{2=22} &= S_{222}(\delta^K_{il}\delta^K_{kj}\delta^K_{mn}) -\frac{1}{3}\mathbf{J}_{222}\\
    \mathbf{J}_{222=} &= S_{222}(\delta^K_{in}\delta^K_{kl}\delta^K_{mj}) -\frac{1}{3}\mathbf{J}_{222}\\
    \mathbf{J}_{2-2-2-} &= S_{222}(\delta^K_{ik}\delta^K_{lm}\delta^K_{nj}) - \frac{1}{9}\mathbf{J}_{222} - \frac{1}{3}\left( \mathbf{J}_{22=2}+ \mathbf{J}_{2=22}+\mathbf{J}_{222=}\right).
\end{align}

\section{Shape Convergence}
\label{sec:appendix_convergence}

\begin{figure*}
    \centering
    \includegraphics[width=\textwidth]{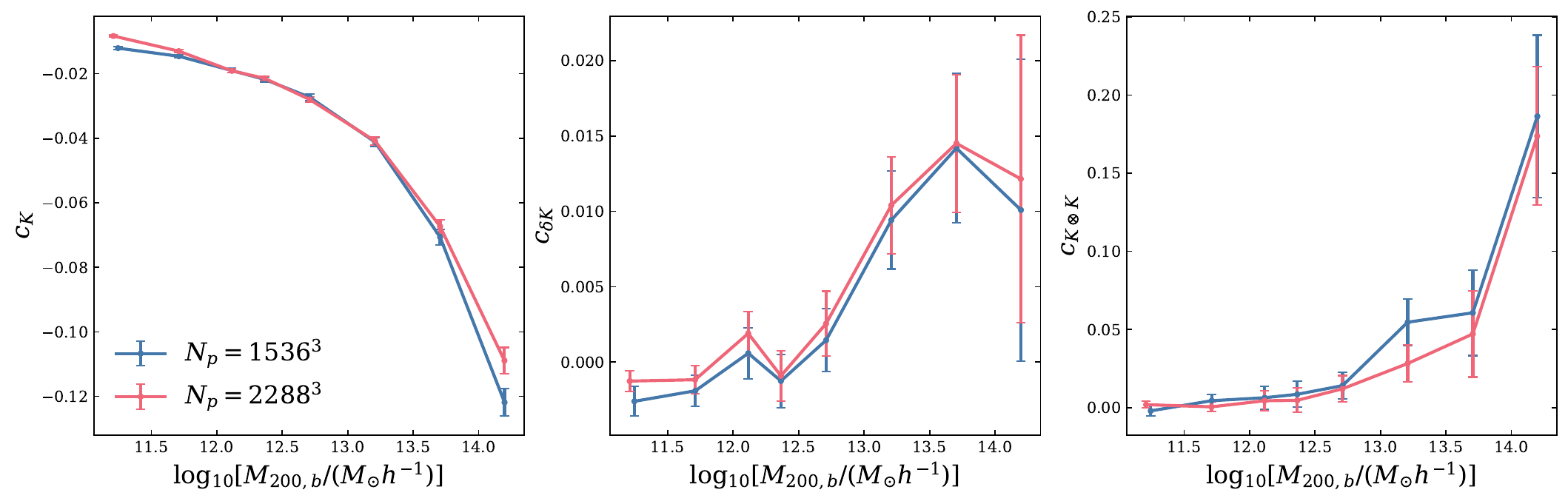}
    \caption{Comparison of measurements of shape-biases $c_K$, $c_{\delta K}$ and $c_{K\otimes K}$ in two versions of the Narya simulation, with increasing resolution. For these measurements, shapes were computed using the FoF halo definition. One can see that the results generally match within the error bars, except for measurements of $c_K$ below $M_{200,b}=10^{12}M_\odot h^{-1}$. The damping scale employed in measuring these bias parameters is of $k_d=0.1h$Mpc$^{-1}$ }
    \label{fig:convergence_test}
\end{figure*}

In this appendix, we display measurements of the shape-bias parameters obtained from two simulations, both run with Narya cosmology, but at different resolutions. This test allows us to define a convergence criterion, which is used throughout the entire article. As can be seen from Figure (\ref{fig:convergence_test}), these bias parameters are in agreement with each other for $M_h>10^{12}M_\odot h^{-1}$. This mass corresponds to halos with roughly 270 particles in the low-resolution Narya simulation, so we define our convergence criterion to be that halos must have more than 300 particles.

\section{Comparison to Akitsu}
Similar measurements to the ones reported in this article were performed by \cite{Akitsu_2023}, but for the Eulerian bias parameters. Nevertheless, in that work, they present expressions for translating the bias parameters from Eulerian to Lagrangian space. Therefore, when comparing our results to theirs, we always transform our measurements according to Equation (2.15) of \cite{Akitsu_2023}.

\end{appendix}

\end{document}